\documentclass[twocolumn]{aastex631}
\usepackage{multirow}
\usepackage{caption}
\usepackage{gensymb}
\usepackage{graphicx}
\usepackage{amsmath,amssymb}
\usepackage{parskip}
\usepackage[shortlabels]{enumitem}
\usepackage{float}
\usepackage{subcaption}
\usepackage{comment}
\usepackage{booktabs}
\usepackage{rotating}
\usepackage{natbib}
\usepackage{hyperref}

\begin{document}
\title{Anatomy of the Star-formation in a Tidally Disturbed Disk galaxy: NGC\,3718}
\author{Chandan Watts}
\affiliation{Indian Institute of Astrophysics,
              II Block, Koramangala, Bengaluru 560 034, INDIA. \\}
\affiliation{Pondicherry University, R.V. Nagar, Kalapet, 605014, Puducherry, India}
\correspondingauthor{Chandan Watts}
\email{chandan@iiap.res.in, chandanwatts510@gmail.com}

\author{Mousumi Das}
\affiliation{Indian Institute of Astrophysics,
              II Block, Koramangala, Bengaluru 560 034, INDIA. \\}
\author{Sudhanshu Barway}
\affiliation{Indian Institute of Astrophysics,
              II Block, Koramangala, Bengaluru 560 034, INDIA. \\}


\begin{abstract}
We present a UV, optical and near-infrared (near-IR) study of the star-forming complexes in the nearby peculiar galaxy NGC\,3718, using UVIT, GALEX, Spitzer and DECaLS imaging data. The galaxy has a disturbed optical morphology due to the multiple tidal arms, the warped disk and the prominent curved dust lanes, but in the near-IR, it appears to be a bulge-dominated galaxy. Its disturbed morphology makes it an excellent case to study star formation in a tidally disturbed galaxy that may have undergone a recent minor merger. To study the distribution and properties of the star-forming clumps (SFCs), we divided the galaxy within the R$_{25}$ (B band) radius into three parts -- the upper, central and lower regions. Using the UV band images, we investigated the warped star-forming disk, the extended tidal arms, and the distribution and sizes of the 182 SFCs. Their distribution is 49, 60 and 73 in the galaxy's upper, central and lower regions, respectively. We determined the UV color, star-formation rates (SFRs), star-formation density ($\Sigma_{SFR}$) and ages of the SFCs.The central disk of the galaxy has a larger mean $\Sigma_{SFR}$ which is $\sim$3.3 and $\sim$1.6 times higher than the upper and lower regions, respectively. We also find that the  SFCs in the central disk are older than those in the tidal arms. Our study thus shows that minor mergers can trigger the inside-out growth of galaxy disks, where the younger SFCs are in outer tidal arms and not in the inner disk. 
\end{abstract}
\setlength{\parindent}{20pt}
\setlength{\parskip}{0pt}
\section{Introduction} \label{sec:intro}
Galaxy interactions produce distinct morphological features around galaxies in the form of extended tidal arms and rings of stars and gas \citep{yadav.etal.2023}. The interactions can be of two types, the first is the distant interactions that occur during the early stages of galaxy mergers or during flybys when the two galaxies are still separate but in close proximity \citep{10.1046/j.1365-8711.2000.03083.x,2008ApJ...681..232L}. The second type consists of galaxies that are either just about to merge or have recently done so \citep{1996MNRAS.279L..47A,Conselice_2003,10.1111/j.1365-2966.2008.13531.x,2009ApJ...697.1971J}. The galaxy mergers can be major (mass$-$ratios $\geq$ 0.25), minor (0.1 $\leq$ mass$-$ratios $<$ 0.25) or mini mergers (0.01 $\leq$ mass$-$ratios $<$ 0.1). \citet{bottrell2023illustristng}, compared the roles of mergers and found that mini-mergers are responsible for 55\% of merger-driven star formation and 70\% of merger-driven asymmetric structures around galaxies for z $\leq$ 0.7.

Before merging, the galaxies move in many orbits around each other, leading to the formation of extended but often distorted tidal rings around the galaxies \citep{barnes.hernquist.1992}. Interaction between  galaxies can also cause significant gas transfer between their disks, resulting in features such as polar-rings or double rings \citep{bournaud.combes.2003}. Good examples are the Helix galaxy, the Annulus, the cartwheel galaxy and the Narrow ring galaxies \citep{1998ApJ...499..635B}. During interactions, significant amounts of stellar and gas mass pulled out in the tidal tails , which are prominent at optical wavelengths, can have different morphology and often show substructures, such as those observed in NGC 646 \citep{yadav.etal.2023}. How are these tidal arms shaped? It depends upon the orbit of the intruder galaxy, the inclination and sense of direction i.e. retrograde or prograde orbits, as well as the relative masses of the galaxies \citep{kumar.etal.2021}. 

Galaxy mergers can trigger star formation in galaxies, especially if the galaxies are gas rich. Hence star formation is an important tracer of galaxy mergers, especially in gas rich minor mergers where the gas pulled out from the galaxies is often associated with tidal arm star formation.  According to simulations, star formation is caused by the tidal interactions between the merging galaxies, which results in enhanced cloud collisions and shocked gas. Then the hot, shocked gas cools to form stars \citep{2009PASJ...61..481S,2004MNRAS.350..798B}. Hence, the star formation is often observed as a star-forming knots along the tidal arms. Simulations show that the interactions also cause gas inflows towards the centers of the interacting galaxies resulting in central star formation or sometimes starbursts (e.g.\citet{2009ApJ...706...67R,2015MNRAS.446.2038R, 2013MNRAS.430.1901H, 2020MNRAS.492.2075B, 2015MNRAS.448.1107M, 2019MNRAS.485.1320M, 2021MNRAS.503.3113M, 2020MNRAS.498.2323J}).

Multiwavelenth studies of galaxy mergers from UV to radio wavelengths can provide valuable insights into the dynamics of mergers, the star formation, and how it contributes to galaxy growth. The most effective probe of star-formation is UV because it is due to emission from massive O and B-type stars, in the UV wavelength range of 120-180 nm (FUV) and 180-300 nm (NUV). The FUV emission arises from the hot O, B stars and the NUV emission generally comes from slightly cooler O,B, and A type stars \citep{koda.etal.2012}.  So FUV emission traces star forming regions of  age 0-100 Myr and NUV emission traces star formation ages of 0-200 Myr \citep{Kennicutt_2012}. The UV color (FUV-NUV) can give an idea of the age of the star-forming regions.

NGC\,3718 is a good example of a tidally disturbed galaxy whose morphology suggests it has undergone a recent merger \citep{2015A&A...580A..11M}. It has a  prominent dust lane that is associated with a warped disk and is clearly visible in the color composite image, as well as multiple tidal arms as shown in Figure \ref{fig:label arms}. It is a nearby galaxy with redshift z=0.00331 and is at a distance of 14.2 Mpc (Table \ref{tab:intro-table}). It has a LINER L1.9 AGN at its centre \citep{2006A&A...455..773V}. NGC\,3718 has a companion galaxy, NGC 3729, which is a starburst galaxy and may have had a past interaction with NGC\,3718. This maybe the reason why NGC\,3718 is tidally disturbed. However, we did not find any connection between the two galaxies, such as a tidal bridge or tail. \citet{1985A&A...142..273S} found a very weak feature of asymmetric hydrogen distribution of  NGC 3729 towards the NGC\,3718, but it is not enough to prove the connection between them. These galaxies belong to the loose Ursa major group, and they are separated from each other by a distance of $\sim$ 47.93 kpc.

Some researchers classified NGC\,3718 as a nearly polar ring galaxy \citep{2009AJ....137.3976S} due to the large tilt of one of its tidal arms with respect to the galaxy plane, while others have classified the galaxy as SB(s)a pec \citep{1991rc3..book.....D}, after noting the presence of a dust lane connected to a tidal arm. \citet{1985A&A...142..273S} compared the galaxy with spindle galaxies due to its ring-like features and also fitted a tilted ring model to illustrate the projection effects in a more clear manner. The author also found that the angle between the rings and the disc increases rapidly at low radii and reaches at large radii asymptotically an angle of almost 90\textdegree. However, its true morphology is still not clear \citep{2015A&A...580A..11M}.

There have been many studies done on NGC\,3718, which clearly show that it has been undergoing some interaction in the past but none of the studies could find any connection with a companion galaxy \citep{2004A&A...415...27P, 2005A&A...442..479K}. \citet{2009AJ....137.3976S}, found that diffuse spiral arms are visible in the blue band and new stars have formed in the twisted neutral hydrogen (HI) gas layer. They found that the HI gas extends far beyond the stellar body and does not share the stellar kinematics. \citet{2004bdmh.confE.109J} also mentioned that NGC\,3718 displays tidal streams resulting from tidal interactions and the analysis done by \citet{2015A&A...580A..11M} strongly indicate that it is a merger remnant, as it shows a combination of spiral photometry and elliptical-like kinematical properties.

In this paper, we use UV data to understand the star formation in NGC\,3718, with a focus on the star formation in the tidally distorted arms. In Section 2, we will discuss the data and its reduction. In section 3, we cover the analysis done using the UV, IR data and the approximate estimation of the age of the disk and tidal arms. Section 4 describes the results obtained from the data and its connection with the previous studies done on NGC\,3718. Section 5 will summarize the results. Throughout the paper we use the flat $\Lambda$CDM model with $H_0$= $70 km/s/Mpc$, $\Omega_{m} = 0.286$ and $\Omega_{vac} = 0.714$ which corresponds to a cosmological scale of 0.068 $kpc/{\arcsec}$ for NGC\,3718.

\section{Data} \label{sec:data}
To study star formation in NGC\,3718, we used data in the UV wave band as UV traces star formation for longer periods compared to H$\alpha$ emission \citep{Kennicutt_2012}. We used archival data from the Ultraviolet Imaging telescope (UVIT) which is mounted on the AstroSat Satellite telescope\footnote{\url{https://astrobrowse.issdc.gov.in/astro_archive/archive/Home.jsp}}.The UVIT observes in the FUV (130-180 nm), NUV (200-300 nm) and visible (320-550 nm) bands. It has a spatial resolution of $\sim$1.4" (FWHM) in FUV and $\sim$1.2" (FWHM) in NUV with a FOV (field of view) of 28'. We used UVIT archival data in the N245M and F148W bands of NGC\,3718 and CCDLAB (version:v17) to reduce the Level 1 raw data (OBS-DATE= '2016-05-29'). CCDLAB is a UVIT image reduction pipeline that works in an automated mode and does geometric correction, drift correction and astrometry on the data \citep{2021JApA...42...30P}. CCDLAB was utilized to correct geometric distortion, bias and dark frame subtraction, background noise subtraction, flat-field illumination, and spacecraft drift to create images for each orbit. The orbit-wise images were merged to create science-ready images (Level 2 data) to increase the signal-to-noise ratio. To provide the world coordinate system (WCS) information to the science-ready images, we have used the "AstroQuery" menu in CCDLAB, which utilizes the same region file from Gaia DR3 to provide the WCS solution to the image.

We also used Galaxy Evolution Explorer (GALEX) archival data in both FUV and NUV bands\footnote{\url{ http://dx.doi.org/10.17909/93cq-qj93}}. GALEX is a NASA space telescope that works in both Imaging and spectroscopy modes of FUV (1344-1786 \AA) and NUV (1771-2831 \AA ) wavebands with a FOV of 1.2\textdegree. It has a spatial resolution of $\sim$4.2" (FWHM) in FUV and $\sim$5.3" (FWHM) in NUV with a pixel scale of 1.5"/pixel. We used the science ready data provided in the MAST archive. We utilized the FUV and NUV background subtracted (counts/sec) data which has been already reduced using the GALEX pipeline available at the MAST archive for photometric analysis (OBS-DATE= '2006-02-25'), and specifications of the data are mentioned in table \ref{tab:DATA} 

To study the old stellar population and dust distribution, we used the Spitzer (3.6 $\mu$m) archival data of the S4G\footnote{\url{https://doi.org/10.26131/irsa425}} (Spitzer Survey of Stellar Structure in Galaxies) survey which consists of 2352 galaxies mapped with IRAC (Infrared Array Camera) channel 1 (3.6 $\mu$m) and 2 (4.5 $\mu$m) on Spitzer Space telescope \citep{2010PASP..122.1397S}. We have used the science-ready data (OBS-DATE = '2011-02-15') of IRAC channel 1 (3.6 $\mu$m) having FWHM of $\sim$1.7" with a pixel scale of 0.75"/pixel. It was reduced using the S4G pipeline1, and was part of the warm mission.

We also used $g$ and $r$ band DR9 images from DECaLS(The DECam Legacy Survey), which are available on the legacy survey site as coadded images in fits format \citep{2019AJ....157..168D}. The DECaLS survey covers an extragalactic sky area around $\sim$ 19,000 deg$^2$ visible from the northern hemisphere. The DECam (Dark Energy Camera) on the Blanco 4m telescope provides the photometry in the $g$,$r$ and $z$-bands with sensitivity from 400-1000 nm with a pixel scale of 0.262"/pixel. It covers the northern and southern regions with Dec $\le$ 32\textdegree and 34\textdegree respectively. In the next section, we will discuss the analysis done using these data sets.

\begin{table}
\centering
\caption{Physical properties of the galaxy NGC\,3718. The UV diameter is measured using the ellipse fitting method, and the length of the tidal arms are calculated using the cubic spline fitting method.}
\hspace*{-1.7 cm}
\resizebox{1.3\columnwidth}{!}{%
\begin{tabular}{@{}ccc@{}}
\toprule \hline
\multicolumn{3}{c}{Details of galaxy NGC\,3718}  
\\ \midrule
RA                           & 11h32m34.853s      & \multirow{2}{*}{\citep{2007A&A...464..553K}} \\
Dec                          & +53d04m04.518s      &                                       \\
Luminosity Distance          & 14.20 Mpc           &                               \\
Absolute Magnitude (UV)      & -16.30 $\pm$ 0.30     & {\citep{2012AAS...21934001S}}              \\
Absolute Magnitude (Visible) & -21.7               & {\citep{1996AJ....112.2471T}}                      \\
R25  (in B-band)             & 2.338\arcmin ( 9.60 kpc) & {\citep{2014A&A...570A..13M}}                   \\
UV Diameter                  & 51.84 kpc           & \multirow{6}{*}{Figure \ref{fig:label arms}}                     \\
Length of Arm 1                       & 21.01 $\pm$ 0.365  kpc       &                                       \\
Length of Arm 2                        & 50.79 $\pm$ 3.59  kpc        &                                       \\
Length of Arm 3                        & 23.09 $\pm$ 0.389  kpc    &                                       \\
Length of Arm 4                        & 22.24 $\pm$ 2.32  kpc      &                                       \\
Length of Arm 5                        & 46.27 $\pm$ 0.33 kpc     &                                       \\
Length of Arm 6                        & 49.02 $\pm$ 0.391  kpc   &                                       \\ \bottomrule \hline \hline\\
\end{tabular}%
}
\label{tab:intro-table}
\end{table}

\section{Analysis} \label{sec:analysis}
\begin{figure*}[t]
\hspace*{-2.5 cm}
     \centering
     \begin{subfigure}[b]{0.23\textwidth}
     \hspace*{1.8 cm}
         \includegraphics[width=1.2\textwidth]{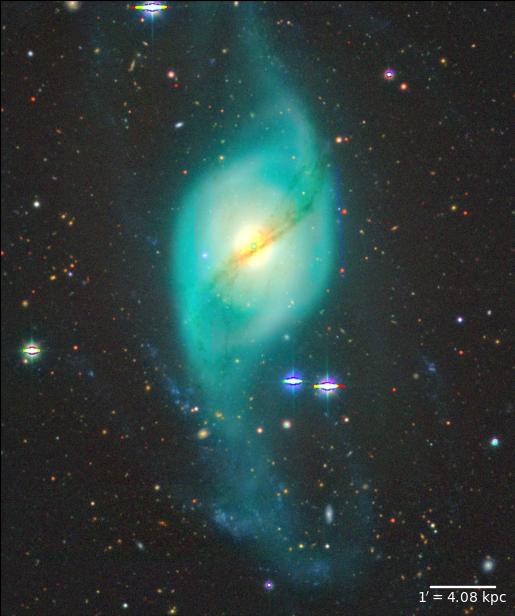}
          \label{fig:decals}
     \end{subfigure}
     \begin{subfigure}[b]{0.42\textwidth}
     \hspace*{2.5 cm}
         \includegraphics[width=1.05\textwidth]{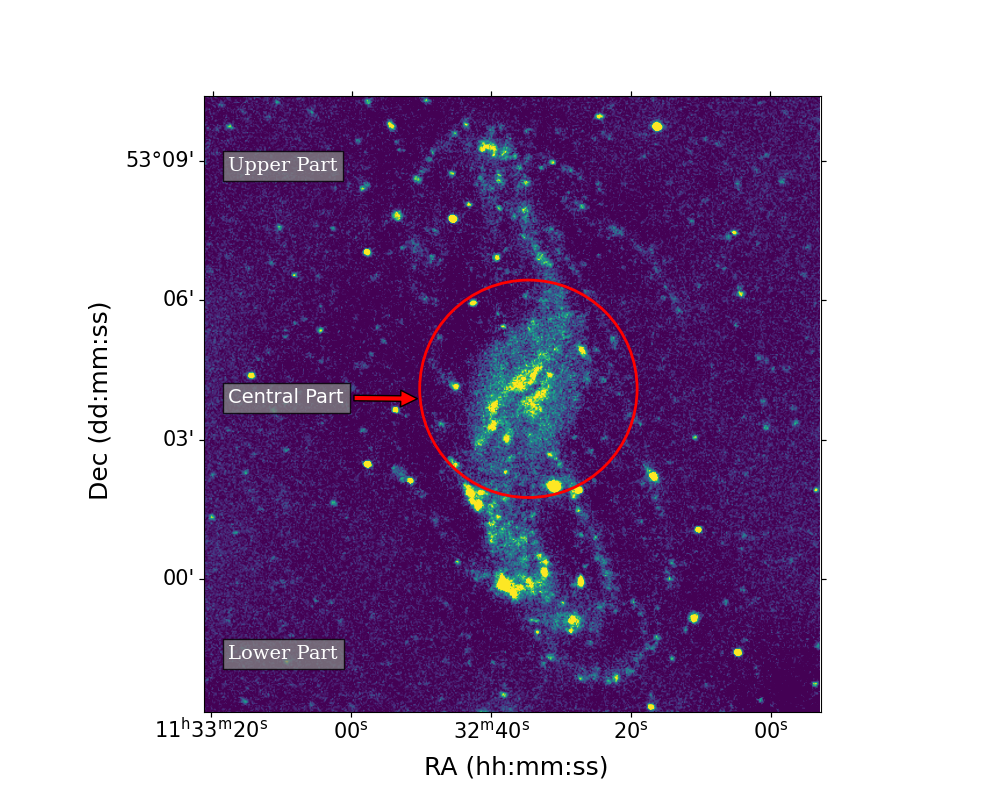}
         \label{fig:Galex NuV image}
     \end{subfigure}
     \begin{subfigure}[b]{0.4\textwidth}
     \hspace*{1.0 cm}
         \includegraphics[width=1.1\textwidth]{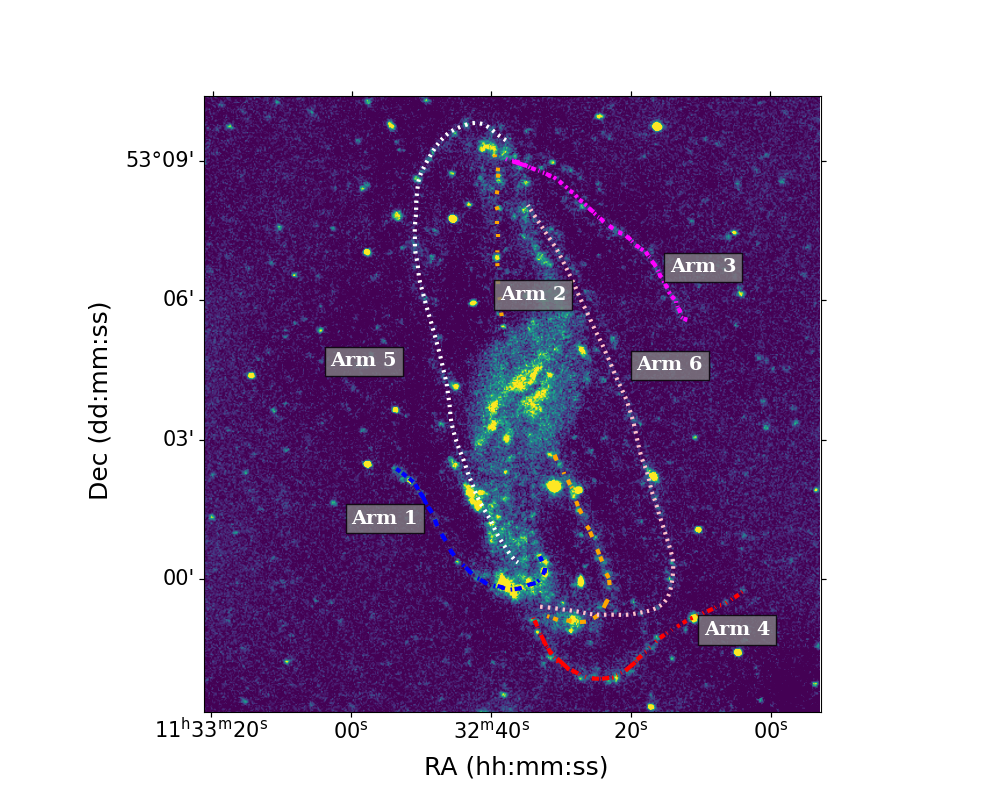}
         \label{fig:arms}
     \end{subfigure}
        \caption{Left: DECaLS grz color composite image of NGC\,3718. Middle: GALEX NUV image with three parts of galaxy: Central Part (R$_{25}$ in B-band), Upper part (Region above the central part) and Lower Part (Region below the central part). Right: The labeling of the tidal arms on the GALEX NUV image. The length of the arms is calculated using the cubic spline fitting method. }
        \label{fig:label arms}
\end{figure*}

NGC\,3718 has many tidal arms, which suggests that it may have undergone a recent merger or experienced several fly-by interactions in the recent past. These unique tidal features can be well studied using multi-wavelength observations. To study the tidal features as well as the warped disk in  detail, we divided the galaxy into three parts as shown in Figure \ref{fig:label arms}. The first part is central part includes the disk lying within the $R_{25}$(B-band) radius, the second is the upper part and the third is lower part lying below the centre of NGC\,3718. We adopted the $R_{25}$(radial location of the isophotes of the galaxy where surface brightness is 25 $mag/arcsec^{2}$) value in the B-band from Hyperleda, which is $2.338^{\prime}$ and performed the analysis on the GALEX data. We have used the GALEX data because the archival UVIT data (Table \ref{tab:DATA}) has a much lower exposure time, and it is not enough to analyse the faint arms of the galaxy. Despite the longer exposure time of the UVIT N245M filter compared to both the GALEX and UVIT F148W filters, our primary focus is on the tidal arms. Although the faint tidal arms are visible in the UVIT N245M filter, they are not detectable in the UVIT F148M filter due to poor signal-to-noise ratio. However, the UVIT images have very good spatial resolution, and the data has enough signal-to-noise (S/N) in the disk region to study the warped disk. So in our study, we used GALEX data to analyze star formation rates and UVIT data to study dust lane patterns in the disk region. We will discuss this in the next section.
\begin{table}[t]
\centering
\caption{Specifications of UVIT and GALEX data. Parameters: Zeropoint (ZP) magnitude and Unit conversion (UC) factor for UVIT and GALEX are taken from  Calibrations done by \cite{2017AJ....154..128T} and \cite{2007ApJS..173..682M}, respectively. The UC is the conversion factor for the filter to convert the counts per sec into flux with units $ erg/sec/cm^{2}/{\AA}$. The signal$-$to$-$noise ratio (SNR) is calculated for individual clumps.}
\hspace*{-2.5 cm}
\resizebox{1.3\columnwidth}{!}{%
\begin{tabular}{@{}cclllclllcc@{}}
\toprule \hline
\textbf{\begin{tabular}[c]{@{}c@{}}Instrument$\rightarrow$\\ Specifications$\downarrow$\end{tabular}} & \multicolumn{8}{c}{\textbf{UVIT}} & \multicolumn{2}{c}{\textbf{GALEX}} \\ \midrule
\multirow{2}{*}{\textbf{Filters}} & \multicolumn{4}{c}{\multirow{2}{*}{F148W}} & \multicolumn{4}{c}{\multirow{2}{*}{N245M}} & \multirow{2}{*}{FUV} & \multirow{2}{*}{NUV} \\
 & \multicolumn{4}{c}{} & \multicolumn{4}{c}{} &  &  \\
\begin{tabular}[c]{@{}c@{}}\textbf{Exposure Time}\\ (in sec)\end{tabular} & \multicolumn{4}{c}{1102.424} & \multicolumn{4}{c}{2175.442} & 1620.55 & 1620.55 \\
\begin{tabular}[c]{@{}c@{}}\textbf{Unit-Conversion} \\ ( $ \times 10^{-15}$)\end{tabular} & \multicolumn{4}{c}{3.09} & \multicolumn{4}{c}{0.725} & 1.40 & 0.206 \\
\textbf{Zero-Point} & \multicolumn{4}{c}{18.016} & \multicolumn{4}{c}{18.50} & 18.82 & 20.08 \\
\textbf{Signal-to-noise ratio} & \multicolumn{4}{c}{2.33-15.66} & \multicolumn{4}{c}{2.87$-$26.98} & 4.82-51.807 & 4.44$-$107.37 \\ \bottomrule \hline \hline
\end{tabular}%
}
\label{tab:DATA}
\end{table}

\subsection{Measurement of tidal arms} \label{sec:tidalarms}
In Figure \ref{fig:label arms}, we  compare the optical (left) and GALEX NUV (middle) images of the galaxy. The UV data traces the tidal arms very clearly and can be best seen in the GALEX NUV image. Based on visualisation, we marked the x and y positions on these arms and detected six arms, as shown in Figure \ref{fig:label arms}(Right). These arms have a distorted and curved structure, so we used the cubic spline method to calculate their length. Using the x and y positions, we employed the cubic spline method to fit the points and measure the projected length of the arms in order to understand their extent and structure. Additionally, we have calculated the projected UV diameter of the galaxy, for which we have drawn the ellipse on the galaxy in such a way that it covers the entire galaxy. Then, we measured the diameter, which is two times the semi-major axis of the ellipse mentioned in Table \ref{tab:intro-table}
\subsection{Detection of star forming clumps (SFCs)}  \label{sec:clumps}
For the detection of SFCs in FUV , we used the Python module of the source extractor (sep version(1.2.1))\citep{Barbary2016,1996A&AS..117..393B} which finds SFCs above the given minimum threshold depending upon the FWHM of the filters of the telescope. We execute the source extractor on the GALEX FUV and NUV background subtracted images for the clumps detection, parameters used for the source extractor are $detection$\_$threshold = 3.5$ $\sigma$, $deblend$\_$nthresh=32$, $minarea=7$, $deblend$\_$cont=0.0001$. Here, we used a detection threshold of 3.5 $\sigma$ with the aim of extracting the clumps on faint tidal arms. Other parameters involved are: \textit{deblend$\_$nthreshold} specifies the number of thresholds needed for the deblending process, \textit{minarea} corresponds to the minimum equivalent to an area of a circle whose area matches the FWHM of the filter, and \textit{debelend$\_$cont} is the minimum separation between the clumps to overcome the contamination of flux from them. We obtained 257 clumps on the FUV background subtracted image, after instrumental correction and correcting for overlapping between SFCs. The source extractor provides us with the photometric parameters such as semi-major axis (a), semi-minor axis(b), clumps position(x,y in pixels), position angle($\theta$) and Kron radius for the individual clumps. The Kron radius provides the aperture for the individual clumps, and it can cover 90\% of flux around that clump. To check, whether it covers the whole star-forming region, we did a visual inspection for these elliptical aperture sizes so that they do not overlap and exceed the size of the clumps. This kron radius is used to scale the semi-major axis and the semi-minor axis to cover the whole star-forming clump and is used to calculate the flux for all clumps. To check the NUV emission from individual clumps, we calculated the Kron radius in the NUV background subtracted image and got 189 clumps that have the same Kron radius as the FUV-detected clumps. 

\begin{figure}[!h]
\vspace{-2 mm}
         \hspace*{-1 cm}
         \centering
         \includegraphics[width=0.58\textwidth]{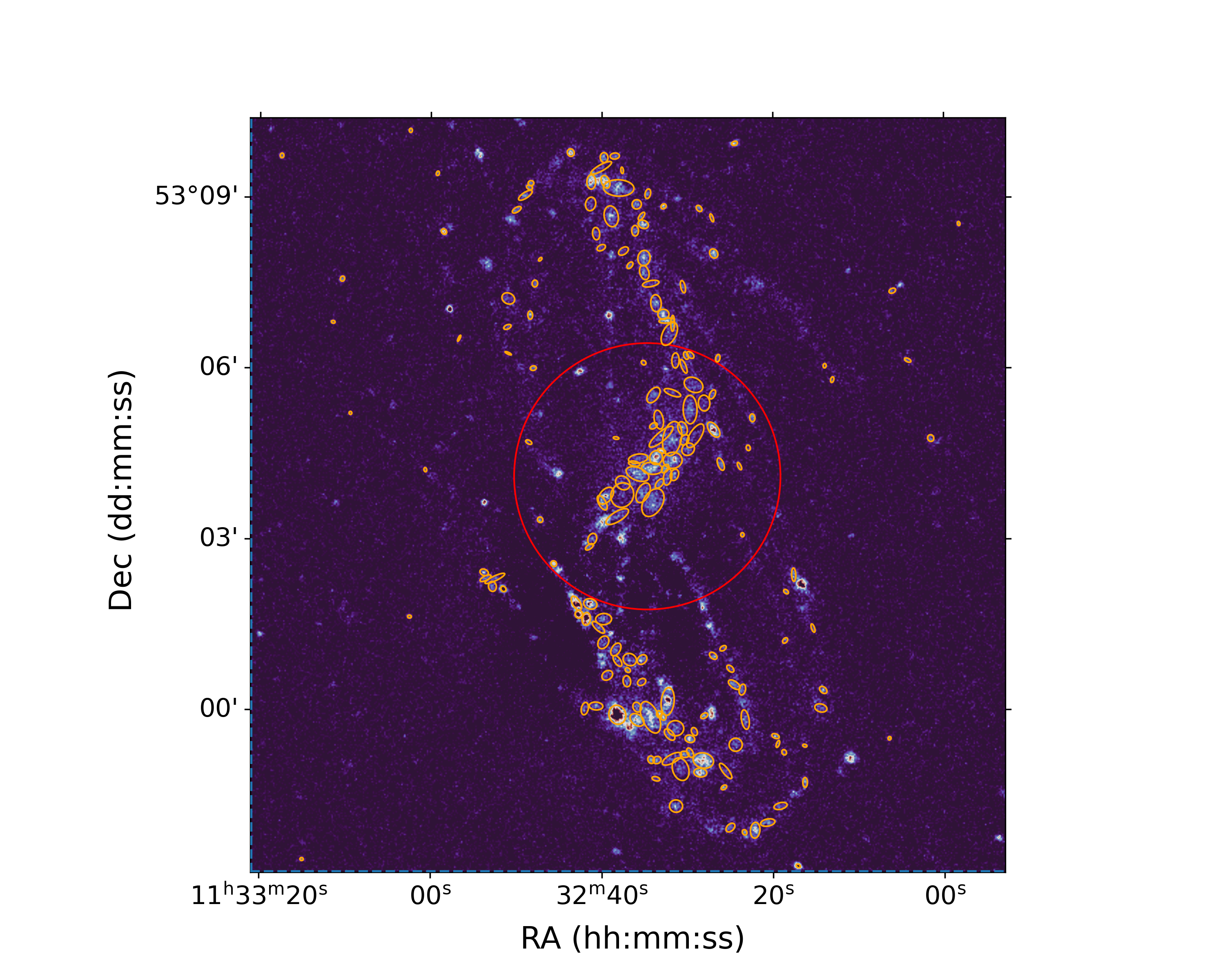}
         \caption{The image above shows the SFCs (orange color) detected in the GALEX FUV and NUV images are superimposed on the FUV image. The red circle is the $R_{25}$ radius in the B-band and includes most of  the central stellar disk.}
         \label{fig:GseNUV}
\end{figure}

The next step is to remove the clumps that are not part of the galaxy by matching the clumps with the simbad catalogues.For this purpose, we utilized the SAOImageDS9 (version 8.5). In SAOImageDS9, within the analysis menu, we used the Simbad database in the catalogs option. We have already converted clumps into a DS9 region file and cross-matched them with the Simbad database for that region. In this way, we obtained 182 clumps\footnote{A Table with parameters of 182 clumps can be provided upon request to the first author.} that are part of NGC\,3718 as shown in Figure \ref{fig:GseNUV}. Using the parameters given by source extractor, we calculated the area of the elliptical aperture and the fluxes of the 182 clumps. As discussed above, we divided the galaxy into three parts, we found that there are more clumps detected in the lower part than in the other parts. The number of clumps detected in the upper, central and lower parts are 49, 60 and 73 respectively. 

\subsection{Calculation of Extinction corrected magnitude}  \label{sec:extinction}
The FUV and NUV emission from the individual clumps can provide us with an idea of the age, the formation history of the tidal tails and the warped disk. It can be obtained using the UV color-magnitude diagram, i.e. the (FUV-NUV) vs FUV diagram in this case. The calculation of (FUV-NUV) color is done using the calibration parameters \citep{2007ApJS..173..682M} such as zero-point(ZP) and unit conversion factor(UC) and the relation $m_{AB} = -2.5log(CPS) + ZP$. As we know, the magnitude we are calculating using the above relation, is affected by galactic extinction. To calculate the extinction corrected magnitude we use the mean R$_V$-dependent extinction law which is, \citep{1989ApJ...345..245C}
\begin{equation} \label{equ:RV}
    \langle\frac{A(\lambda)}{A(V)}\rangle = a(x) + \frac{b(x)}{R_V} 
\end{equation}
where A(V) and A$(\lambda)$ are the extinction in the visual band and in the desired wavelength band respectively. A(V) = 0.039 \citep{2011ApJ...737..103S}, a(x) and b(x) is the wavelength-dependent parameters and are taken from \citet{1989ApJ...345..245C}, by considering the FUV and NUV wavelengths. Here, x is the inverse of the wavelength ($\lambda^{-1}$), R$_V$ is 3.1 for the milky way. Using the value of A$(\lambda)$, we obtained the milky way extinction correction in the magnitudes obtained from the individual clumps.

\subsection{Color-magnitude map}  \label{sec:color map}
\begin{figure*}[!htb]
     \centering
     \begin{subfigure}[b]{0.49\textwidth}
         \centering
         \includegraphics[width=\textwidth]{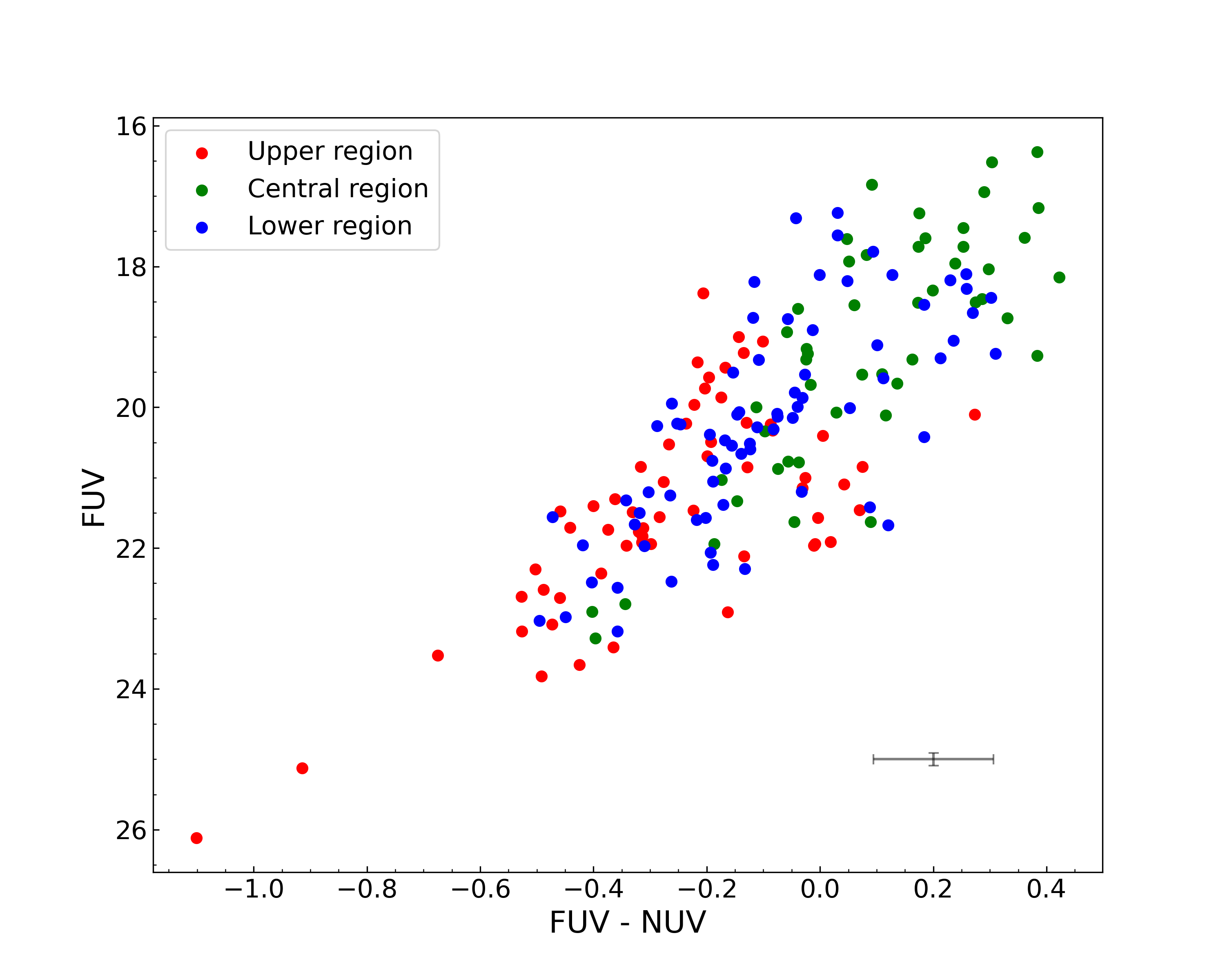}
     \end{subfigure}
     \begin{subfigure}[b]{0.49\textwidth}
         \centering
         \includegraphics[width=\textwidth]{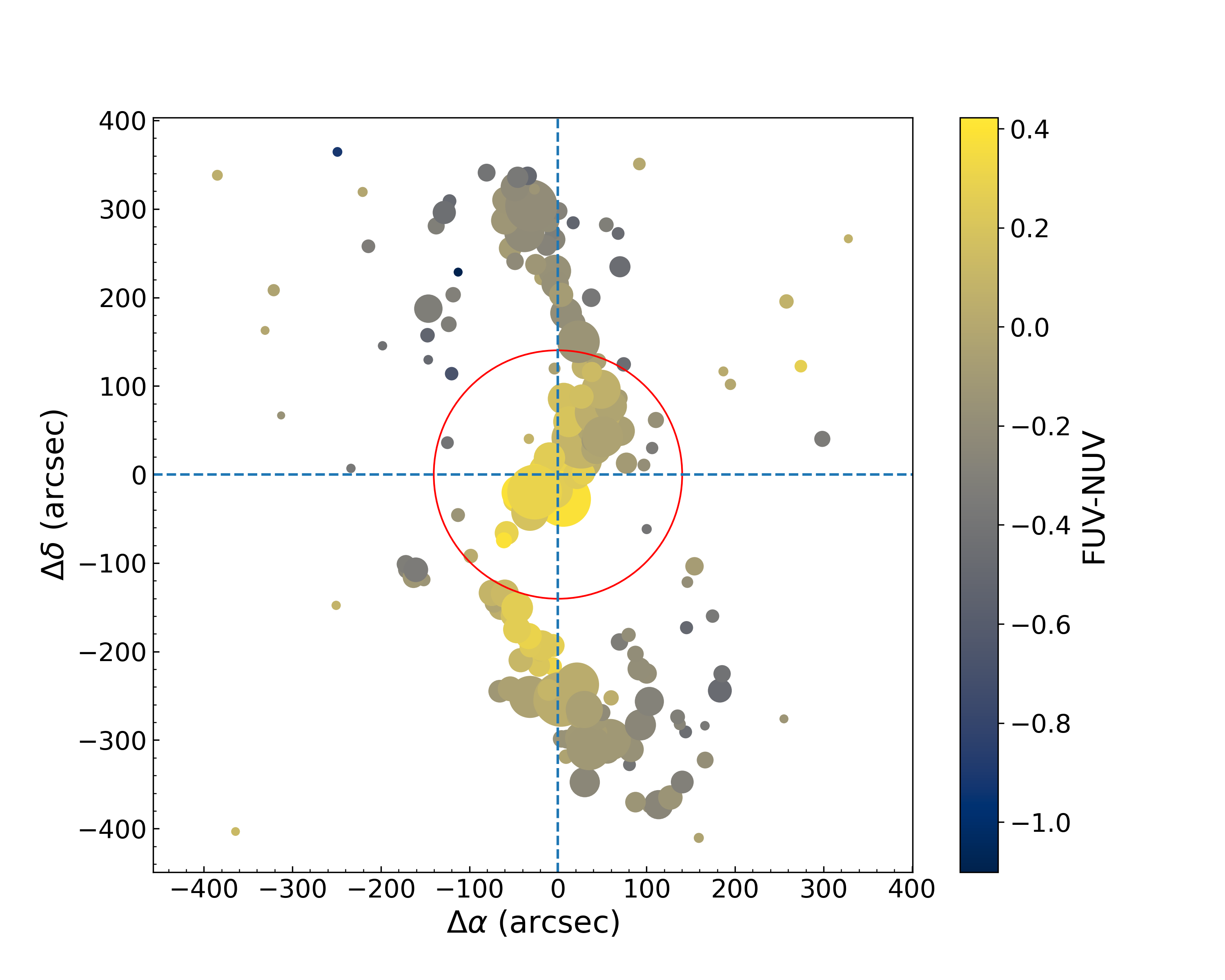}
     \end{subfigure}
        \caption{The variation of the (FUV-NUV) color in the warped disk and tidal arms of NGC\,3718. Left: The FUV-NUV vs FUV comparison in the warped disk and tidal arms. This shows that the central part of the galaxy has emissions from an old stellar population. Right: The variation of FUV-NUV in the galaxy i.e. color-map for the whole galaxy. Here, We have taken the size of the clumps as equal to half of the area.}
        \label{fig:GColor-map}
\end{figure*}

\begin{figure*}[!htb]
     \centering
     \begin{subfigure}[b]{0.253\textwidth}
     \hspace{-1.5 cm}
         \centering
         \includegraphics[width=\textwidth]{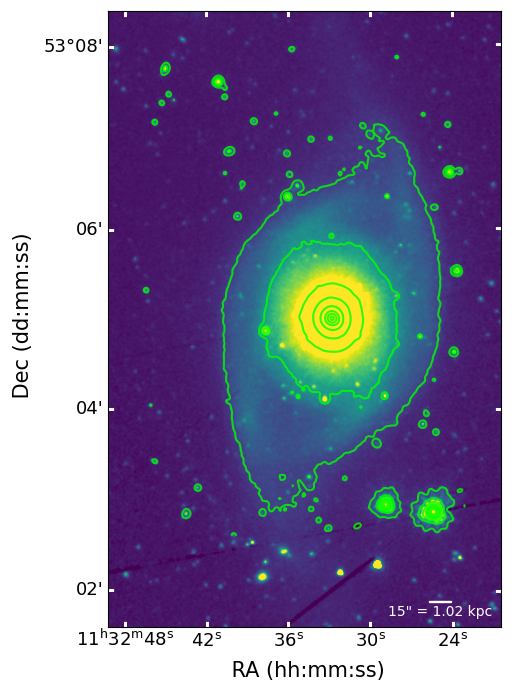}
     \end{subfigure}
     \begin{subfigure}[b]{0.359\textwidth}
     \hspace{-1.0 cm}
         \centering
         \includegraphics[width=\textwidth]{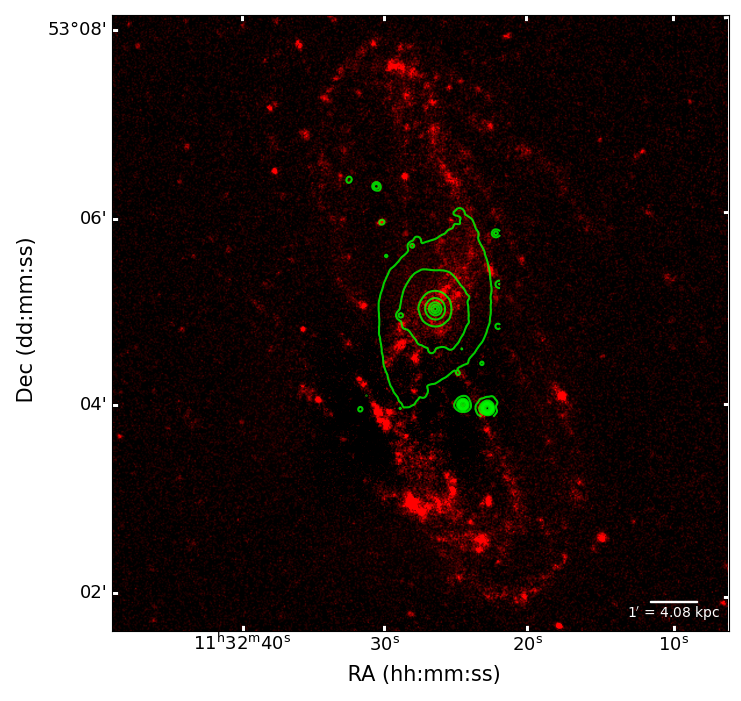}
     \end{subfigure}
     \begin{subfigure}[b]{0.363\textwidth}
     \hspace{-0.3 cm}
     \centering
         \includegraphics[width=\textwidth]{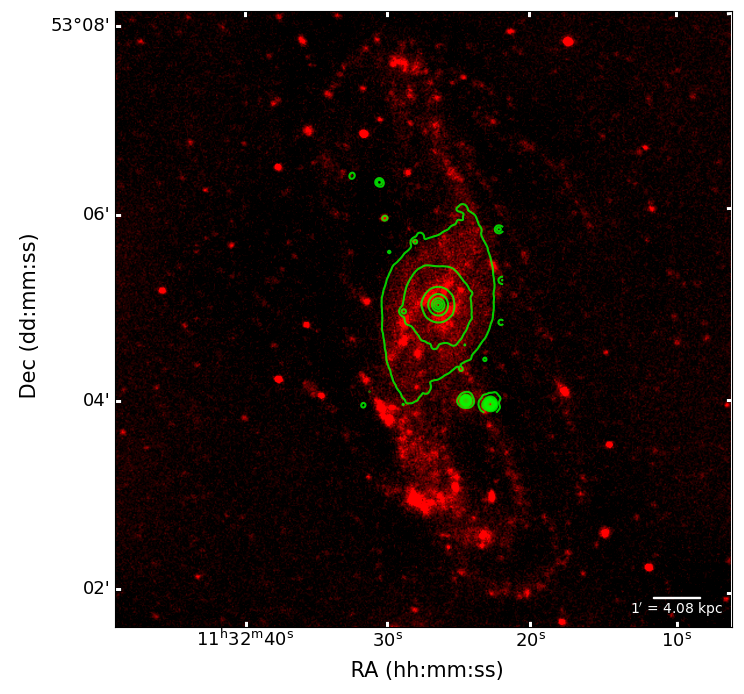}
     \end{subfigure}
        \caption{Left: The contour map of spitzer 3.6 $\mu$m emission showing the prominent bulge. These contours are above 3$\sigma$ noise and levels are for intensity values of $0.02, 0.15, 0.42, 1.01, 2.28, 5.01, 10.90, 23.60, 50.95, 109.89$ MJy/sr. Middle: The overlap of spitzer 3.6 $\mu$m contours on the GALEX FUV image. Right: The overlap of spitzer 3.6 $\mu$m contours on the GALEX NUV image.}
        \label{fig:GContours of phot1}
\end{figure*}

With the help of extinction-corrected magnitudes, we prepared the color-magnitude diagram as shown in Figure \ref{fig:GColor-map}. It shows a variation in FUV magnitude  $\sim$ 16 - 26  with positive and negative color variations after milky way and internal extinction corrections, we discuss this later in section \ref{sec:sfr}. The positive color indicates more NUV emission as compared to FUV emission in the clumps. To see this in a more elaborate way for the warped disk and tidal arms, we can plot the SFC color distribution over the galaxy as shown in the Figure \ref{fig:GColor-map}. The disk part has a positive color, i.e. more NUV emission in the disk compared to the tidal arms. The NUV emission lasts for $\sim$ 0 - 200 Myr and represents an older population in the disk. 
The Spitzer 3.6$\mu$m emission is sensitive to older and cooler stars (K and M-type stars), including red giants and AGB stars with an age of few Gyr. These stars emit strongly in the infrared region. This data is suitable for mapping the older stellar population, particularly in the galactic bulge and disk. On the other hand, the FUV and NUV emission traces the younger and hotter population (A and B-type stars). By overlapping both datasets, we want to confirm two things. First, we want to compare the distribution of younger and older stellar populations. Second, we want to compare the tidal arms at different wavelengths. To check the distribution of different stellar populations, we used the Spitzer  3.6 $\mu$m data that gives us an idea of the older stellar population, we prepared contour maps of the Spitzer data above 3$\sigma$ noise and overplotted these on the FUV and NUV data as shown in Figure \ref{fig:GContours of phot1}. It clearly shows that the warped disk has an older stellar population than the tidal arms, which can be due to the interaction of NGC\,3718 with the merging galaxy. This results in the redistribution of stellar material and gas over the tidal arms. Surprisingly, we didn't find any older stellar populations around the tidal arms, confirming the FUV-NUV results that only a hot younger population is associated with the tidal arms.\\

\subsection{Prominent Dust lane }
NGC\,3718 has a prominent dust lane in the centre of the disk which is associated with the warped disk. The FUV and NUV SFCs are clearly distributed around the dust lane, and basically follow the warped. We used both FUV and NUV UVIT detected SFCs, as the UVIT has a better spatial resolution than GALEX, and clearly shows the distribution. 

\begin{figure}[t]
\hspace*{-1 cm}
         \centering
         \includegraphics[width=0.6\textwidth]{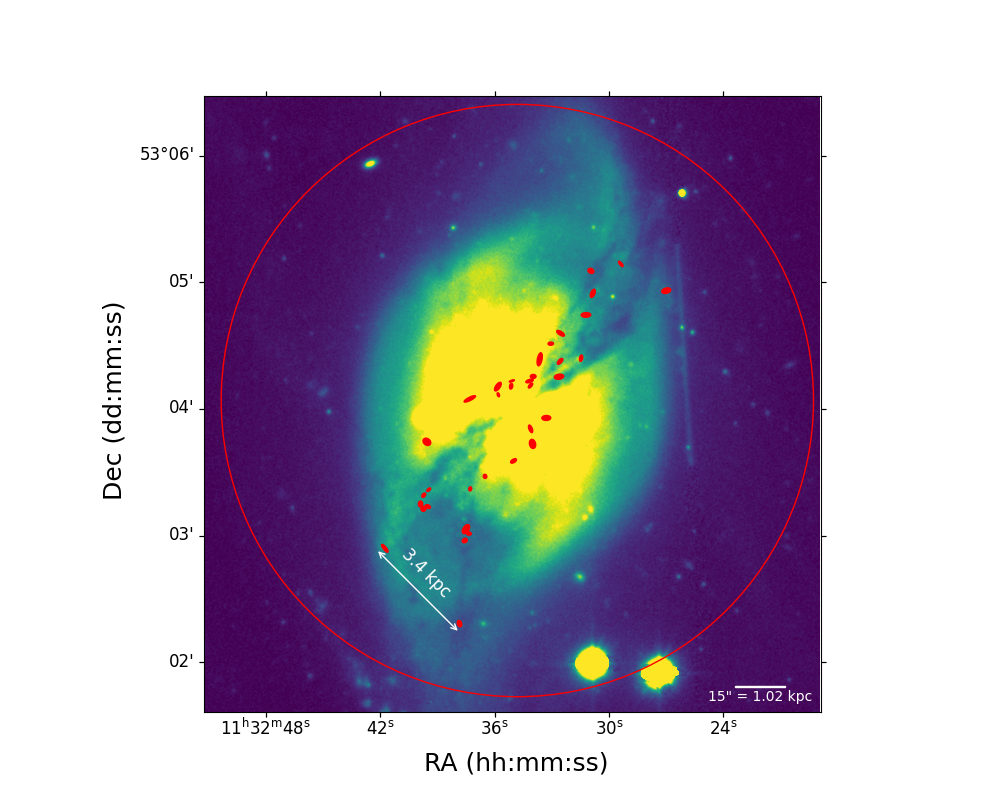}
         \caption{The UVIT disk clumps (red color) overlaid on optical g-band image. }
         \label{fig:dustlane}
\end{figure}

We have overlaid the UVIT clumps on the optical g-band image as shown in Figure \ref{fig:dustlane}. The UVIT NUV image (N245M filter) shows the SFC distribution around the dust lane more clearly than other bands and we discuss this later in Section 5.2. We found that the SFCs also follow the twisted path of the warped disk or dust lane, but have a varying width. We measured the scale of the dust lane using UVIT SFCs and the maximum height normal to the disk is computed to be  $\sim$3.4 kpc in the southern part of the dust lane. This is fairly large compared to most star-forming disks that have height $<$200 pc. It is probably due to the warped nature of the disk which may produce more star formation at higher disk heights. \\
\section{Results}

\subsection{FUV and NUV images}
The FUV and NUV images of NGC 3178 provides a detailed view of the star formation in the the warped disk, the tidal arms, and the dust lane. The UV follows the dust lane associated with the disk very closely, which suggests that there is considerable star formation associated with the disk but the UV emission is extincted by the dust. A closer look at the UV emission around the dust lane reveals a gap in the lower part of the galaxy disk, and faint tidal arms associated with SFCs arise from the disk ends (Figure \ref{fig:GseNUV}). The gap in the warped disk maybe due to the presence of outflows associated with the disk star formation or AGN activity, or it maybe due to the pericenter passage of a companion galaxy during the interaction. Furthermore, there are no SFCs detected in the gap. Also, a careful examination of the UV images reveals that the  tidal arms of the galaxy are not entirely smooth and are broken in some areas, indicating that they may not be only due to an interaction with a companion galaxy - a merger event and star formation activity may also be important. 

However, the 3.6 $\mu$m Spitzer image of the galaxy shows only a large bulge which is associated with an older stellar population, the warped disk and tidal arms are surprisingly not visible at all (Figure \ref{fig:GContours of phot1}). The disk is probably over shadowed by the near-IR luminosity of the bulge. The stellar density in the tidal arms is low and hence not visible in near-IR. Both the disk and tidal arms are dominated by a younger stellar population. Therefore, UV analysis is necessary to obtain a complete picture of the interaction history and the associated star formation in the galaxy, as it is absent in the longer wavelength images.

\subsection{Age and metallicity} \label{sec:age}
\begin{figure*}[t]
     \centering
     \begin{subfigure}[b]{0.49\textwidth}
         \includegraphics[width=\textwidth]{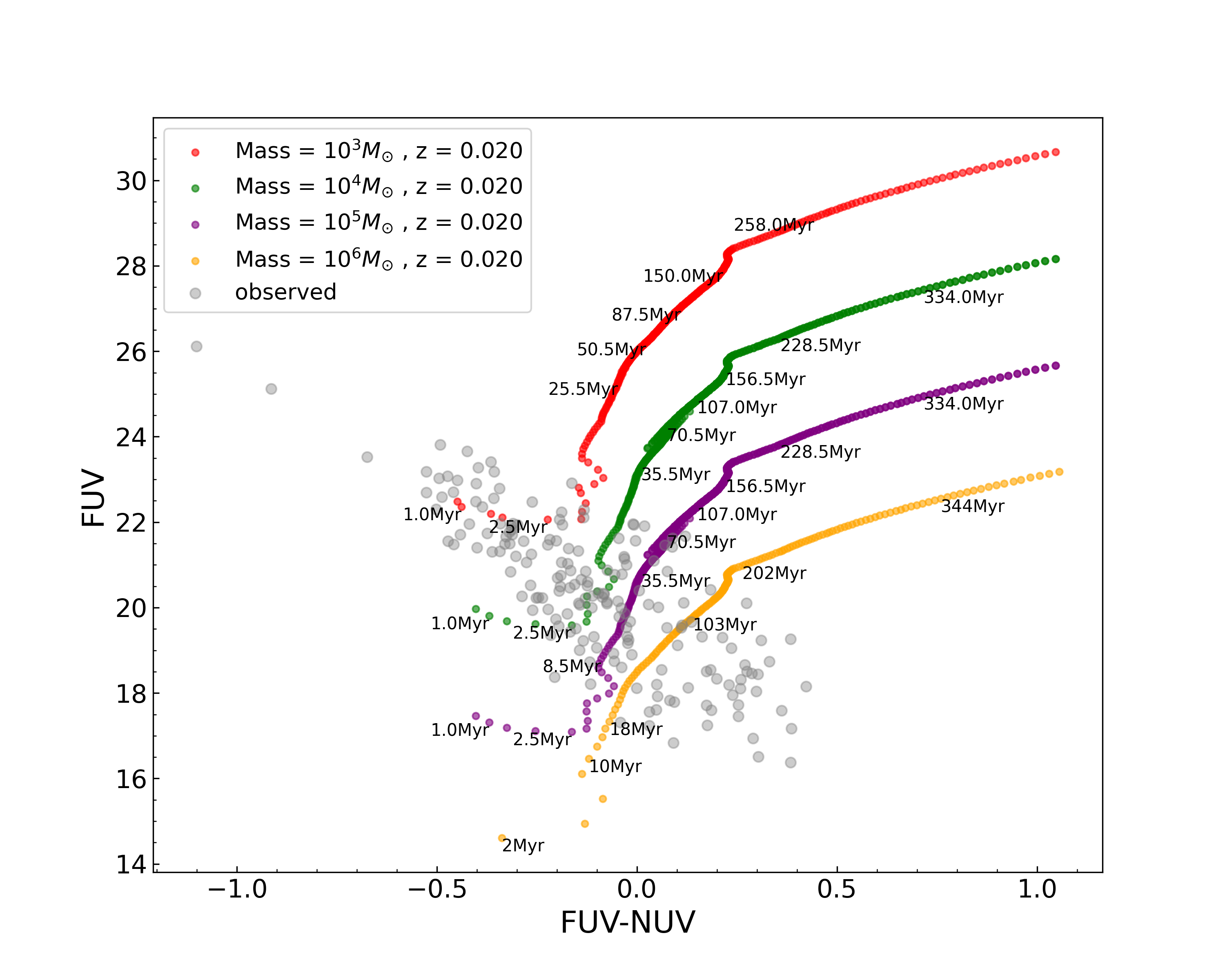}
     \end{subfigure}
     \begin{subfigure}[b]{0.49\textwidth}
          \centering
         \includegraphics[width=\textwidth]{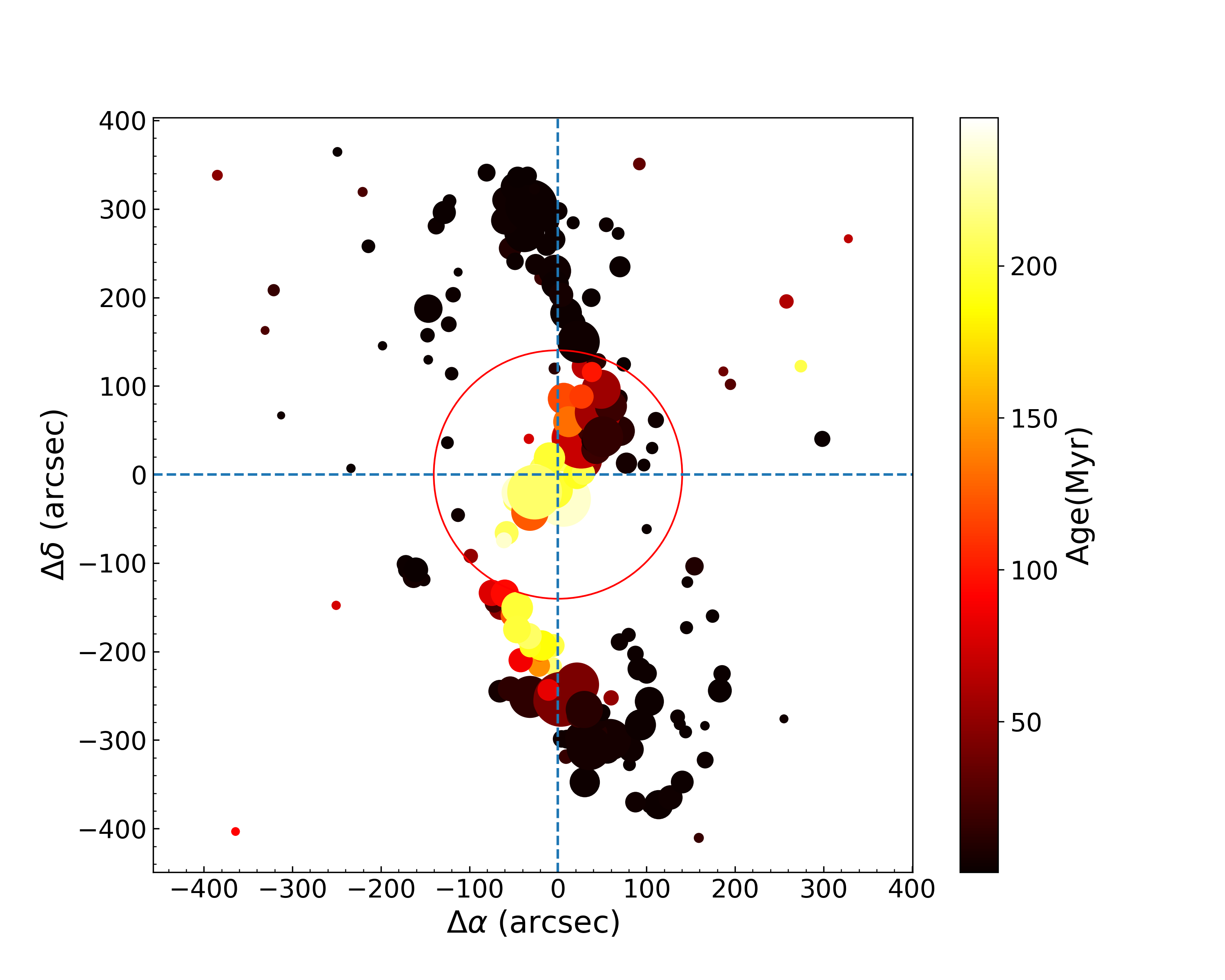}
     \end{subfigure}
    \caption{Left: Comparison of starburst99 models and observed color. Right: Age distribution in whole galaxy with clumps having metallicity, z = 0.020 and mass = $10^5 M_{\odot} $.} 
        \label{fig:Gs99}
\end{figure*}

For the calculation of age and metallicity of the tidal arms and warped disk, we used starburst99 \citep{1999ApJS..123....3L}. We used models with SFC masses of $ 10^3, 10^4, 10^5 $and $10^6 M_{\odot} $, the Kroupa initial mass function (IMF) and metallicity with the Padova track ( z = 0.020, i.e. solar metallicity). We convolved the model fluxes with the GALEX filter effective area to obtain the fluxes corresponding to the filter wavelength. These fluxes are used to calculate the FUV and NUV magnitudes. From these models, we prepared the (FUV-NUV) vs FUV tracks for the models and compared it with the observed data, in order to estimate the clump masses (assuming solar metallicity) in the central disk and tidal arms (upper and lower parts) as shown in the Figure \ref{fig:Gs99}.

Among all the tracks, only the one with z = 0.020 and $10^{5} M_{\odot}$ covers the FUV magnitude range of the observed clumps the best. Therefore, we used it for the age calculation. We linearly interpolated the age of the clumps by comparing them with the model to obtain an idea of the age distribution. All these measurements of color-magnitude using starburst99 show that the central disk part has an older stellar population and tidal arms are younger than the warped disk (Figure \ref{fig:Gs99}). This difference in age maybe due to the merger with another galaxy or it maybe due to more than one passage of a companion galaxy in a fly-by interaction.

\subsection{Calculation of the star formation rate} \label{sec:sfr}
\begin{figure*}[t]
     \centering
     \begin{subfigure}[b]{0.49\textwidth}
         \centering
         \includegraphics[width=\textwidth]{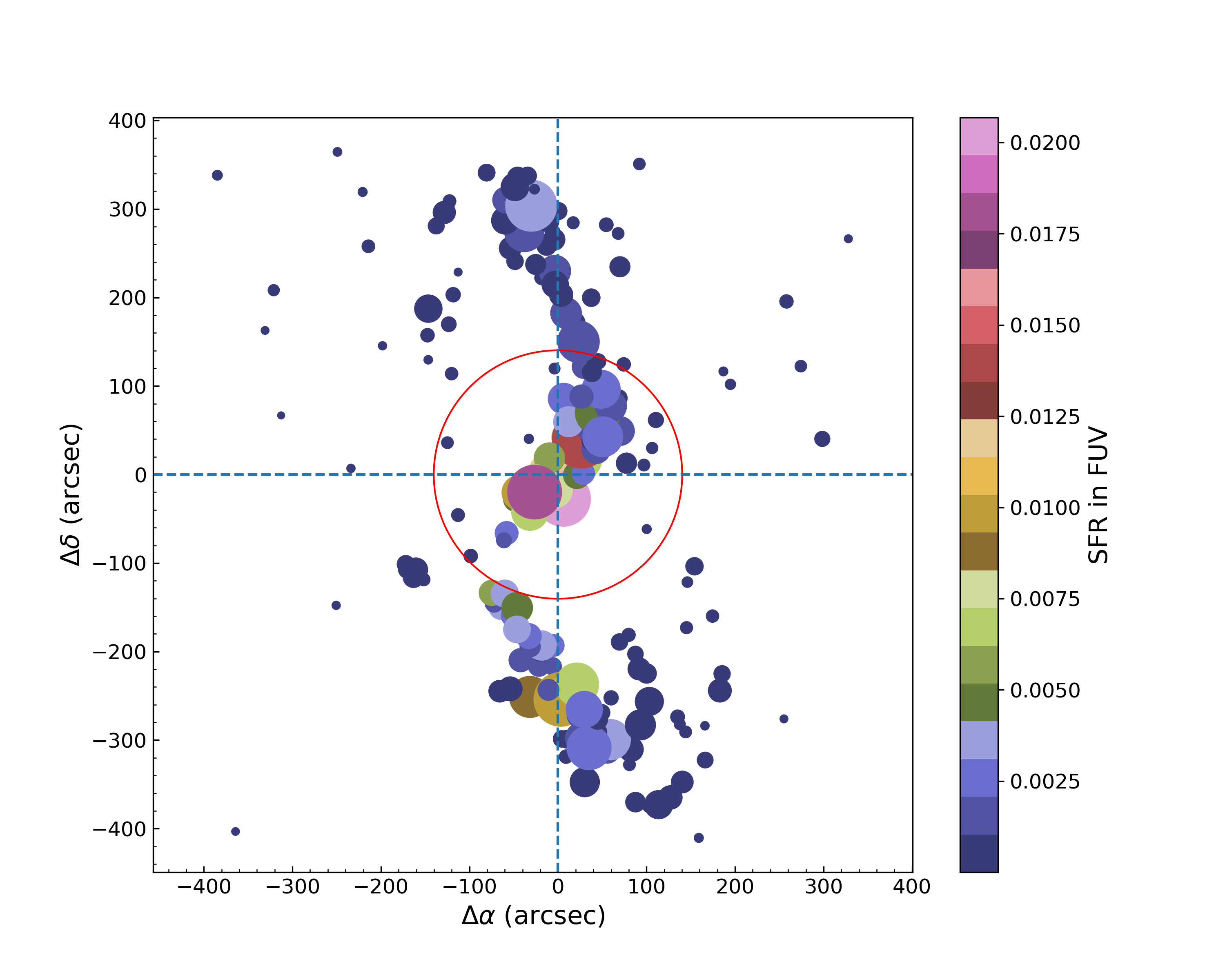}
     \end{subfigure}
     \centering
     \begin{subfigure}[b]{0.49\textwidth}
         \centering
         \includegraphics[width=\textwidth]{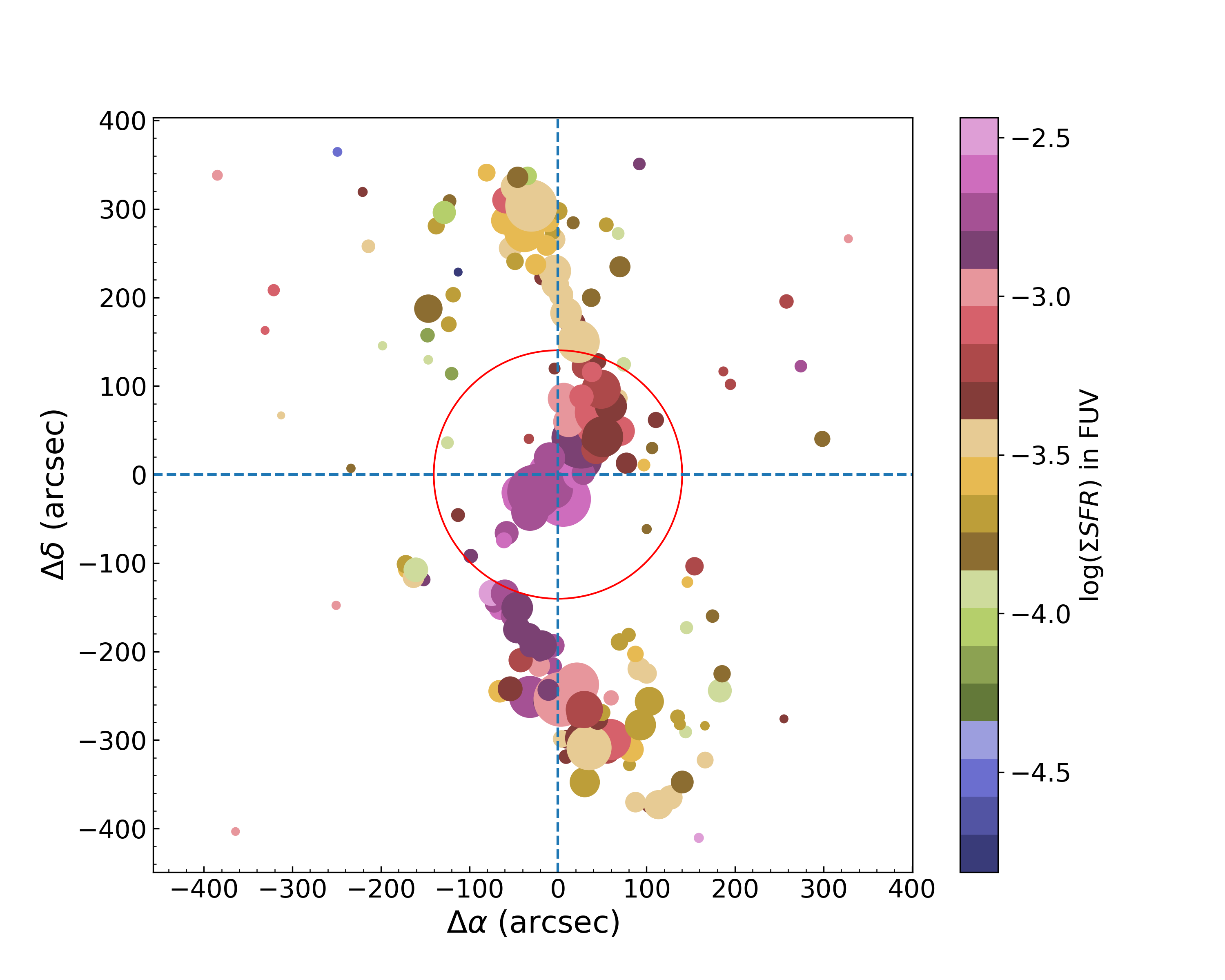}
     \end{subfigure}
        \caption{Left: Star-formation rate in FUV. Right: Star-formation density in FUV. }
        \label{fig:GSFR}
\end{figure*}

One of the main objectives of this paper is to calculate the SFRs of the complexes in the target galaxy with the help of UV data. For the calculation of SFR, we have used the \citet{2013MNRAS.436.3135E} star formation relations,

\begin{equation*} \label{equ:SFR_1}
\begin{split}
    SFR_{FUV} = 4.6 \times 10^{-44}(L_{FUV})_{corr}\\
    SFR_{NUV} = 6.8  \times 10^{-44}(L_{NUV})_{corr}
\end{split}
\end{equation*}
Here, $SFR_{FUV}$ and $SFR_{NUV}$ are star formation rates(in $M_{\odot}/yr$) in FUV and NUV respectively, $(L_{FUV})_{corr} $  and  $(L_{NUV})_{corr}$ are the luminosities (in $erg/s$) calculated after the internal extinction correction. For the internal extinction corrections, we used the $\beta$- UV slope method \citep{2016ApJ...822...42O} which uses the Milky Way extinction correction magnitude for wavelengths in the FUV and NUV as shown below.

\begin{equation*} \label{equ:betaslope}
\begin{split}
     \beta_{UV} = -0.4\frac{m_{FUV} - m_{NUV}}{log(\frac{\lambda_{FUV}}{\lambda_{NUV}})} -2 \\
               = 2.252(m_{FUV} - m_{NUV}) -2
\end{split}
\end{equation*}
This  $\beta$-slope method is used to study the effects of dust extinction on the UV continuum by fitting it with the corrected foreground extinction which was done with the power law fitting in the wavelength range $1250 \leq \lambda \leq 2600 \AA$ as $f(\lambda) \propto {\lambda}^{\beta_{UV}}$. Using this $\beta_{UV}$ slope, we can calculate the color excess E(B-V) \citep{2018ApJ...853...56R} as

\begin{equation*}
    \begin{split}
     \beta_{UV} = -2.616 + 4.684 E(B-V) \\
     E_{s}(B-V) = (0.44 \pm 0.03) E(B-V) \\
     \end{split}
\end{equation*}
Here, $E_{s}(B-V)$ is the color excess of the stellar continuum and is related to the $E(B-V)$ by the above relation and then internal extinction is calculated using the starburst reddening curve $K^{\prime}(\lambda)$ as

\begin{equation*}\label{equ:extinction_correction}
     A_{\lambda} = K^{\prime}(\lambda)E_{s}(B-V) \\
\end{equation*}
\begin{equation*} \label{equ:extinction_correction_1}
    \begin{split}
     K^{\prime}(\lambda) = 2.659(-2.156 + \frac{1.509}{\lambda} -\frac{0.198}{\lambda^2} +\frac{0.011} {\lambda^3}) + R^{\prime}_{v}  \\
     (0.12 \mu m \leq \lambda \le 0.63 \mu m )
    \end{split}
\end{equation*}
Here, $R^{\prime}_{v} = 4.05 \pm 0.80$ \citep{2000ApJ...533..682C}, it is taken such that the reddening curve can recover most of the light lost due to the dust. To observe, how the internal extinction is behaving all over the galaxy, we checked the pattern of the internal extinction for each clump, which shows the mere extinction near the lower part of the disk which is due to the dust lane in the disk and observed in the color-composite DeCALS image (Figure \ref{fig:label arms}).
Using the extinction measurements, we can calculate the extinction-corrected luminosity as

\begin{equation*}
    \begin{split}
     L_{corr} = L_{obs}\times 10^{0.4 E_{s}(B-V)K^{\prime}(\lambda)}
    \end{split}
\end{equation*}
Here, $L_{obs}$ is the observed flux. Using this, we get the $SFR_{FUV}$ and $SFR_{NUV}$ for the FUV and NUV data. We have shown the SFR in FUV in the Figure \ref{fig:GSFR}. 

These plots show clearly that SFR is larger near the central part of the disk compared to the other parts of the galaxy. We find that the SFR varies from $9.828 \times 10^{-7}  - 3.051 \times 10^{-2} M_{\odot}/yr$ with a mean star formation of $2.033 \times 10^{-3} M_{\odot}/yr$ in NUV and $2.615 \times 10^{-6}  - 2.068 \times 10^{-2} M_{\odot}/yr$ with a mean star formation of $1.712 \times 10^{-3} M_{\odot}/yr$ in FUV. We also determined the SFR with respect to the sizes of the clumps, i.e. to determine the correlation of SFR and clump size. To do this we converted the area of the clumps into arcseconds which is obtained using the plate scale of GALEX FUV and NUV filters. The next step is to convert it into physical parameters such as $kpc^{2}$ which can be done by considering the scaling of 1\arcsec=0.068 kpc. This area is used to calculate the star formation density of the clumps, by dividing the SFR (in $M_{\odot}yr^{-1}$) to the area (in $ kpc^2$) as shown in Figure \ref{fig:GSFR}. The star formation density has a mean value of $7.792 \times 10^{-4} M_{\odot}/yr/kpc^{2}$ in FUV and $8.844 \times 10^{-4} M_{\odot}/yr/kpc^{2}$ in NUV. For the different parts of the galaxy, these numbers are mentioned in table \ref{tab:sfr}. From this table, we can clearly see that the central part i.e. the disk of the galaxy, has a larger SFR than the other parts of the galaxy.
\begin{table}[]
\centering
\caption{Mean star-formation rate (SFR) and star-formation density (SFD) for the upper, central and lower part of the galaxy.}
\hspace*{-1 cm}
\resizebox{\columnwidth}{!}{%
\begin{tabular}{@{}c|cc|cc@{}}
\toprule \hline
 & \multicolumn{2}{c}{\textbf{\begin{tabular}[c]{@{}c@{}}SFR \\
 ($\times 10^{-3}$ \textit{$M_{\odot}/yr$})\end{tabular}}}
 & \multicolumn{2}{c}{\textbf{\begin{tabular}[c]{@{}c@{}}SFD \\ 
 ($\times 10^{-3}$ \textit{$M_{\odot}/yr/kpc^{2}$})\end{tabular}}} \\ \midrule
 & \textbf{FUV} & \textbf{NUV} & \textbf{FUV} & \textbf{NUV} \\
\textbf{Upper part} & 0.417 $\pm$ 0.032  & 0.373 $\pm$ 0.023 & 0.373 $\pm$ 0.058 & 0.354 $\pm$ 0.051\\
\textbf{Central part} & 3.777 $\pm$ 1.404 & 4.843 $\pm$ 1.364 & 1.254 $\pm$ 0.352 & 1.578 $\pm$ 0.344\\
\textbf{Lower part} & 1.391 $\pm$ 0.268 & 1.511 $\pm$ 0.216 & 0.793 $\pm$ 0.129 & 0.854 $\pm$ 0.108\\ \bottomrule \hline \hline
\end{tabular}%
}
\label{tab:sfr}
\end{table}

\section{Discussion} \label{sec:discussion}
\subsection{Star formation in tidal arms} 
NGC\,3718 is a very interesting example of a galaxy undergoing tidal interactions or a merger event. Its proximity (D=14.2 Mpc) allows us to study the star formation in the numerous tidal arms, which may have originated due to a minor merger event or a fly-by interaction. We have identified six arms ( Table\ref{tab:intro-table}) based on the GALEX FUV and NUV images. These arms are distorted and broken, indicating that there have been several passages of the minor galaxy before it merged with the disk. 

The tidal arms have numerous SFCs. To check the variation of size of the clumps and how they evolve all over the galaxy, we have prepared a map of NGC\,3718, where the SFCs have sizes proportional to their clump diameters, i.e. two times the semi-major axis of the clumps. We have observed that there is a trend for clumps in the inner galaxy to be larger in size compared to the clumps at the tip of the tidal tails. The diameter of the star-forming clumps varies from 0.43 - 5 kpc with a mean diameter of 1.429$\pm$0.007, 2.167$\pm$0.009 and 1.676$\pm$0.006 kpc in the upper, central and lower parts of the galaxy. In the following paragraphs we discuss the distribution of the clumps arm-wise.

ARM 1: The number of clumps detected in arm 1 are very few ($\sim$ 14) because of its low surface brightness, and they do not fully trace the arm. However, we have detected the clumps at the tip and the starting point of the tail. The diameter decreases with distance away from the galaxy, such as from 2.773 $\pm$ 0.014 kpc to 1.264 $\pm$ 0.047 kpc.

ARM 2: This also has significantly fewer clumps ($\sim$ 21) tracing the arms but is still present at the end points of the arm, as well as two clumps near the galaxy center. The clumps detected near the galaxy center have small sizes ($\sim$0.818 $\pm$ 0.075 kpc), and the other clumps sizes decrease with distance away from the galaxy center, such as 2.084 $\pm$ 0.027 kpc to 1.289 $\pm$ 0.066 kpc.

ARM 3 and ARM 4: A similar trend is observed in these arms where the clump size varies from 1.206 $\pm$ 0.081 kpc to 0.630 $\pm$ 0.054 kpc and 2.085 $\pm$ 0.057 kpc to 0.502 $\pm$ 0.030 kpc for arm 3 and arm 4, respectively.

ARM 5 and ARM 6: Arm 5 and 6 have a half-loop shape; combining both loops will make one ring around the galaxy. We have seen a similar trend here as well, but we found that larger clumps are present in arm 6, which varies in size from $\sim $2-3 kpc. 

All these analyses show of the SFC sizes along the tidal arms appears to fade away as with increasing distance from the center of the host galaxy. It looks like due to the passage of a low mass companion galaxy, it has these asymmetric arm structures. A recent study by \cite{bottrell2023illustristng} also confirms that $\sim$70 $\%$ of the asymmetric structures are formed by mini mergers.

\citet{2015A&A...580A..11M} demonstrate that NGC3718 shows the characteristics of a past merger event, such as the extended tidal arms and the warped disk. Our calculation of the spatial distribution of the SFRs also suggests this. We find that the central disk SFCs have larger SFRs compared to the outer tidal arms. Also, the central disk hosts an older stellar population compared to the outer tidal arms. Therefore, from the age and SFR estimates, it is clear that the tidal arms have a younger stellar population with lower SFRs compared to the central disk. This can be explained by a past minor merger event which caused the inflow of gas to the center of the galaxy as well as in the tidal arms. A larger gas surface density will lead to star formation through gravitational instabilities. But the inner disk has a larger stellar surface density (and hence self gravity) compared to the more diffuse tidal arms. Hence, the SFRs are higher and the star formation proceeds more rapidly in the inner disk compared to the outer tidal arms, which have lower stellar densities and hence gravitational instabilities take longer to set in.

Another possible reason could be the influence of the AGN (LINER) present in the center of the disk. Outflows from AGN can cause the compression of the surrounding gas and trigger star formation in the inner region of the galaxy (AGN feedback). However, LINERs are low luminosity AGN and usually do not show strong outflows or jets \citep{wylezalek.etal.2020}. 

\subsection{Relation of UV emission with CO flux}
\begin{figure}
         \vspace{-9mm}
         \hspace*{-1.7cm} 
         \includegraphics[width=0.65\textwidth]{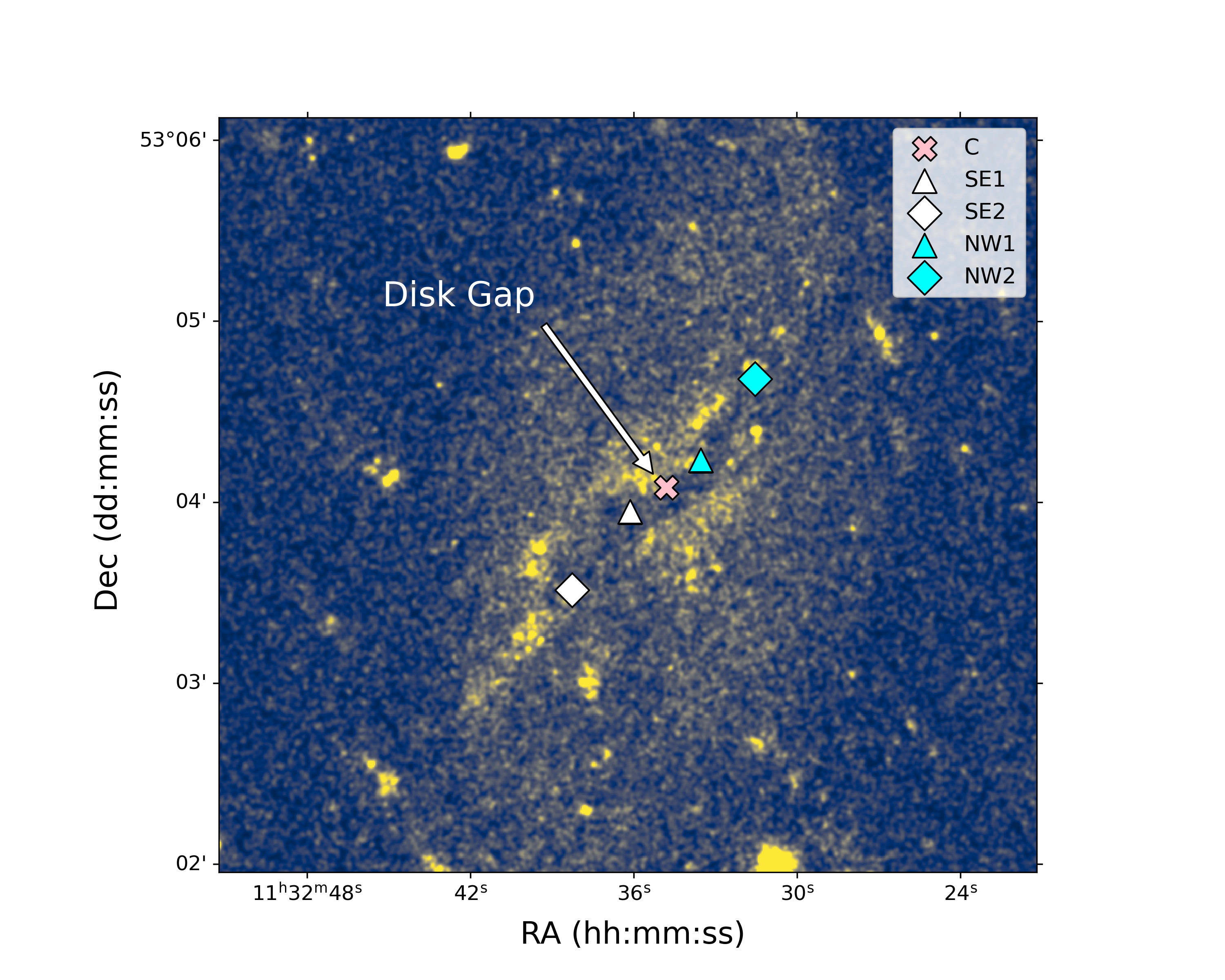}
         \caption{ Relation between UV and CO flux. symbol represents the five peaks central(C), NW1, NW2, SE1 and SE2 of CO(1-0) emission on UVIT-NUV245M image. }
         \label{fig:Gaps}
\end{figure}

Another exciting result from our study of NGC\,3718 is the gap in the dust lane aligned with the disk, which is clearly visible in the GALEX and UVIT NUV images. In Figure \ref{fig:Gaps}, we have marked the disk gap; it appears fairly broad and lies in the lower part of the disk. It could be due to dust trapped inside the gap, and molecular gas could be associated with the dust. To clarify this in a more detailed manner, we need to check the presence of molecular gas, and CO emission is the best tracer of the molecular gas. \citet{2005A&A...442..479K} showed in their paper that CO(1-0) emission appears in the form of an "S-like" shape and has mainly five peaks which are marked as C, NW1, NW2, SE1 and SE2. Their study also mentioned that the eastern part (SE1 and SE2) of the CO(1-0) emission has less flux as compared to the western part (NW1 and NW2). We overlaid the five peaks on the FUV image to see how the CO emission relates to the UV emission for this galaxy. We found that most of the CO emission matches the location of the  gap in the disk, and we have also identified the star-forming clumps around that region, which indicate the merger event.
\subsection{Similarities with Polar-ring galaxies}
Polar ring galaxies (PRGs) are galaxies that have a gaseous ring nearly orthogonal to the plane of the disk. NGC 4650a is a famous PRG that has been studied extensively \citep{bournaud.combes.2003}. In the case of NGC\,3718, if we observe the UV image of this galaxy, we can identify the multiple tidal arms. However, none of these arms are orthogonal to the stellar disk plane. \cite{2009AJ....137.3976S}, proposed a polar ring scenario for this galaxy based on their  analysis of the inner parts of the HI disk, and concluded that the outer orbits have a low inclination. \cite{2015A&A...580A..11M} considered the polar ring morphologies suggested by \citet{2009AJ....137.3976S} and investigated the mechanism of the ring formation in this galaxy. For this, the study considered the two mechanisms proposed by \citeauthor{bournaud.combes.2003}, i.e. merger and accretion, and concluded that the merger scenario was the best fit for this galaxy. 

Our analysis of star formation also suggests that this galaxy has undergone a merger in the past. If we examine Figure \ref{fig:label arms} closely, where we have marked the arms on the UV image, we can see that by combining Arm 5 and Arm 6, we can create a ring around the stellar disk of the galaxy. However, this ring is not perpendicular to the galaxy's disk. By comparing the UV structure of this galaxy with the simulations produced by \cite{1998ApJ...499..635B}, where the concept of the victim and intruder was used, it resembles the initial mass ratio of 0.5 at T=22.0 for the gaseous part of the victim in Figure 14 given in the paper of \citeauthor{1998ApJ...499..635B}. This implies that this is one snapshot of this galaxy's transition into a polar ring galaxy. Later, the rings will stabilize and become a double ring or polar ring, as observed in the simulations. Our analysis suggests that this galaxy is not yet a polar ring galaxy and is in an intermediate state. In fact, it is on its way to becoming a multi-ring or polar ring galaxy.
\section{Summary}
In this study, we have used the UV data from GALEX and UVIT to study in detail the morphology and star-formation activity in the disk and tidal arms of the tidally disturbed galaxy NGC\,3718. We have summarized our work as follows:
\begin{enumerate}
    \item NGC\,3718 is divided into three parts to study the arms and disk parts more efficiently; they are the upper, central, and lower parts. The disk lies within the optical radius $R_{25}$, which covers the central part of the galaxy. The upper and lower parts are defined by taking the central part as a reference.
    \item We have identified 182 clumps in the GALEX FUV and NUV images using a source extractor and after applying  all the corrections. The number of clumps detected in the upper, central and lower parts are 49, 60 and 73 respectively. 
    \item Our UV analysis with FUV-NUV color and age estimation suggests that the central part of the galaxy is older than the upper and lower parts of the tidal arms associated with the galaxy. 
    \item Based on visualisation, we labelled the tidal arms and computed their length using the cubic spline fitting method to understand how long these arms are extended. Additionally, we examined the diameter trend of the clumps within the arms and found that it is consistent across all the arms.
    \item We have also calculated the star formation rate and the star formation density of the SFCs. We observed that the central part of the galaxy has more star formation than the other parts. This suggests that star formation happens more rapidly in the inner part than in the diffuse tidal arms.
    \item If we combine the arms 5 and 6 of the galaxy, it results in a ring structure around it. However, the ring is not orthogonal to the disk. Upon comparing the UV morphology of the galaxy with the simulations, we observed that this galaxy is in an intermediate state of becoming a multi ring or polar ring galaxy.  
\end{enumerate}

\section*{ACKNOWLEDGMENTS}
 The authors thank the anonymous referee for the thorough review of the manuscript and scientific feedback, which significantly improved the content and clarity of the paper.  MD and SB acknowledge the support of the Science and Engineering Research Board (SERB) Core Research Grant CRG/2022/004531 for this research. This publication utilizes the data from UVIT, GALEX, Spitzer and DECaLS. UVIT is a part of the AstroSat mission of the Indian Space Research Organisation (ISRO), archived at the Indian Space Science Data Centre (ISSDC). This research has made use of NASA's Astrophysics Data System (ADS), SIMBAD database and the NASA/IPAC Extragalactic Database (NED). \\ \\
 \\
\textit{Facilities}: Astrosat(UVIT), GALEX, Spitzer, Blanco.
\\
\textit{Softwares}: Topcat \citep{2005ASPC..347...29T}, ds9 \citep{2003ASPC..295..489J}, Source extractor \citep{1996A&AS..117..393B, Barbary2016}, Astropy \citep{2018AJ....156..123A, 2013A&A...558A..33A} , Matplotlib \citep{Hunter:2007}, NumPy  \citep{harris2020array} 

\bibliography{NGC3718}

\end{document}